\documentclass[
  aps,prx,twocolumn,
  superscriptaddress
]{revtex4-1}

\usepackage[utf8]{inputenc}
\usepackage[T1]{fontenc}
\usepackage{lmodern}

\usepackage{amsmath,amsfonts,mathtools,bm}
\usepackage{graphicx}
\usepackage{dcolumn}
\usepackage{overpic}
\usepackage{dsfont}
\usepackage{nicefrac}
\usepackage[normalem]{ulem}
\usepackage{xcolor}
\usepackage{siunitx} 

\definecolor{linkColor}{rgb}{0,0.3,0.7}
\usepackage{hyperref}
\hypersetup{
  colorlinks=true, allcolors=linkColor,
  pdfborder={0 0 0}, pdfencoding=auto
}

\begin{document}

\title[Pattern Formation Beyond Turing]{Pattern Formation Beyond Turing: \\ Physical Principles of Mass-Conserving Reaction–Diffusion Systems}

\author{Erwin Frey}
\email{frey@lmu.de}
\author{Henrik Weyer}
\thanks{Present address: Kavli Institute for Theoretical Physics, University of California Santa Barbara, Santa Barbara, CA 93106, USA}
\affiliation{Arnold Sommerfeld Center for Theoretical Physics, Department of Physics, Ludwig-Maximilians-Universit\"at M\"unchen, Theresienstra\ss e 37, D-80333 Munich, Germany}

\begin{abstract}
Intracellular protein patterns govern essential cellular functions by dynamically redistributing proteins between membrane-bound and cytosolic states, conserving their total numbers. 
This review presents a theoretical framework for understanding such patterns based on mass-conserving reaction–diffusion systems.
The emergence, selection, and evolution of patterns are analyzed in terms of mass redistribution and interface motion, resulting in mesoscale laws of coarsening and wavelength selection.
A geometric phase-space perspective provides a conceptual tool to link local reactive equilibria with global pattern dynamics through conserved mass fluxes.
The Min protein system of \emph{Escherichia coli} provides a paradigmatic example, enabling direct comparison between theory and experiment.
Successive model refinements capture both the robustness of pattern formation and the diversity of dynamic regimes observed \emph{in vivo} and \emph{in vitro}. 
The Min system thus illustrates how to extract predictive, multiscale theory from biochemical detail, providing a foundation for understanding pattern formation in more complex and synthetic systems.
\end{abstract}


\maketitle


\section{INTRODUCTION}
\label{sec:introduction}

The physics of cellular systems poses one of the deepest challenges in modern science. While a comprehensive theoretical framework for cellular life remains elusive, physics—particularly theoretical physics—offers powerful tools to uncover the principles that govern cellular organization, dynamics, and function. A productive strategy in this pursuit is the reduction of biological complexity to minimal, tractable subsystems.
Experimentally, bottom-up reconstitution reconstructs specific cellular functions from a few well-characterized components, enabling precise control of biochemical and physical conditions. 
On the theoretical side, coarse-grained and minimal models—often formulated as reaction--diffusion systems or field theories—isolate essential variables and interactions,
and, thereby identify the underlying physical mechanisms.
Importantly, this approach raises fundamental questions: To what extent can the spatiotemporal organization and dynamics of cellular processes be captured by universal physical principles? Which features of biological organization emerge generically in far-from-equilibrium systems, and which reflect biochemical specificity? How far can minimalist approaches advance a predictive theory of living matter?

\begin{figure*}[thb]
\centering
\includegraphics[]{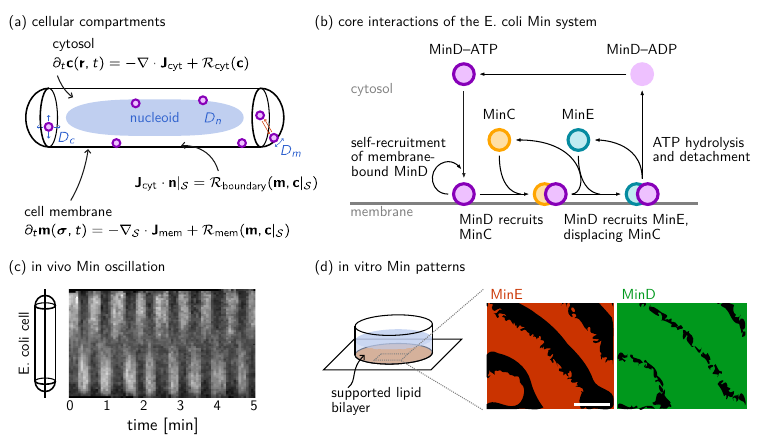}
\caption{
\textbf{Protein pattern formation exemplified by the \textit{E. coli} Min system}. 
(a) Prokaryotic cells comprise the membrane, cytosol, and nucleoid. Membrane-bound proteins diffuse slowly ($D_m$) compared to cytosolic proteins ($D_c$); diffusion in the nucleoid ($D_n$) may differ from that in the cytosol. Biochemical reactions (red arrows) regulate membrane association of proteins (purple). 
(b) The Min system operates via ATP-driven cycling of MinC (yellow), MinD (magenta), and MinE (cyan) between cytosol and membrane. Pattern formation requires only MinD and MinE; MinC inhibits Z-ring formation. 
(c) These interactions generate pole-to-pole oscillations \emph{in vivo}. Shown is a kymograph of MinD fluorescence (white: high intensity; courtesy of Sourjik lab).
(d) \emph{In vitro}, MinD (green) and MinE (red) form dynamic patterns on supported lipid bilayers, including traveling waves (sketch with scale bar: $\sim\SI{50}{\micro m}$)~\cite{Loose.etal2008}.
}
\label{fig:introduction}
\end{figure*}

In this review, we focus on \emph{intracellular protein patterns}, a class of systems where such questions can be addressed in detail. 
Such patterns control essential processes including division, polarity, and intracellular transport by dynamically redistributing proteins between membrane-bound and cytosolic states (Fig.~\ref{fig:introduction}a). 
A common molecular motif is the cyclic switching between inactive and active conformations with different membrane affinities, driven far from equilibrium by nucleotide hydrolysis. 
Because protein production and degradation are negligible on the timescales of pattern formation, these systems are well described by \emph{mass-conserving reaction--diffusion} (McRD) models.  
In contrast to classical Turing-type mechanisms, where patterns arise from the balance of local synthesis and degradation, McRD systems generate patterns through redistribution of the conserved protein mass.

A paradigmatic example is the \emph{Min system} in \emph{E. coli} (Fig.~\ref{fig:introduction}b), where the interactions between just two proteins—MinD and MinE—drive their robust pattern formation (Fig.~\ref{fig:introduction}b). 
\emph{In vivo}, the system produces pole-to-pole oscillation that guides division site placement (Fig.~\ref{fig:introduction}c)~\cite{Raskin.deBoer1999,Hu.Lutkenhaus1999}. 
Remarkably, these dynamics have been reconstituted \emph{in vitro} on supported lipid bilayers from purified components (Fig.~\ref{fig:introduction}d)~\cite{Loose.etal2008}, subsequently revealing a rich diversity of patterns under well-controlled conditions. 
In the following, we offer a biophysical perspective on pattern formation in living systems, highlighting how core principles of mass-conserving reaction–diffusion dynamics connect molecular interactions to mesoscale protein patterns, with the E. coli Min system as a paradigmatic example.

\section{MASS-CONSERVING REACTION--DIFFUSION SYSTEMS}
\label{sec:McRD}

To understand the self-organization of intracellular protein patterns on a mechanistic level, we first turn to the theoretical description of McRD systems. 
These systems provide a general modeling framework for protein dynamics that couple spatial redistribution of proteins via diffusion with local reactions at the membrane–cytosol interface, under the constraint of mass conservation
(Fig.~\ref{fig:introduction}a). 
In what follows, we formulate the governing equations for such systems. 
We then analyze a minimal two-component model that captures the essential physics of these systems and introduces key concepts such as reactive equilibria, mass-redistribution instabilities, and the mass-redistribution potential.

\subsection{General theoretical framework}
\label{sec:McRD_Theory_Framework}

The \emph{spatiotemporal dynamics} of proteins in cells arise from their redistribution between the cytosol and the membrane, described by concentration fields $\mathbf{c}(\mathbf{r}, t)$ in the three-dimensional cytosolic volume and $\mathbf{m}(\boldsymbol{\sigma}, t)$ on the two-dimensional membrane surface $\mathcal{S}$ (Fig.~\ref{fig:introduction}(a)). 
These dynamics are governed by coupled \emph{reaction--diffusion equations} that describe transport and biochemical reactions within each compartment (see Ref.~\cite{Frey.Brauns2022} for a pedagogical introduction).
\begin{subequations}\label{eq:rds}
\begin{align}
    \partial_t \mathbf{c}(\mathbf{r},t)
    &=
    -\nabla \cdot \mathbf{J}_\text{cyt} + {\cal R}_\text{cyt}(\mathbf{c})
    \, ,
    \label{eq:pdes-a}
    \\
    \partial_t \mathbf{m}(\boldsymbol{\sigma},t)
    &= 
    -\nabla_{\mathcal{S}} \cdot \mathbf{J}_\text{mem} 
    + {\cal R}_\text{mem}(\mathbf{m},\mathbf{c}|_{\mathcal{S}})
    \, .
\label{eq:pdes-b}
\end{align}
\end{subequations}
Here, $\mathbf{J}_\text{cyt}$ and $\mathbf{J}_\text{mem}$ denote the cytosolic and membrane fluxes, incorporating both diffusive and possibly advective transport.  
The covariant derivative $\nabla_{\mathcal{S}}$ accounts for the membrane’s curved geometry. 
The terms $\mathcal{R}_\text{cyt}$ and $\mathcal{R}_\text{mem}$ describe local biochemical reactions in the cytosol and on the membrane. 
Notably, the membrane reactions explicitly depend on the cytosolic concentration $\mathbf{c}|_{\mathcal{S}}$ near the membrane, reflecting the biochemical coupling between both compartments, e.g., in the attachment of cytosolic proteins.
This coupling also manifests in the \emph{reactive boundary condition} at the membrane:
${\mathbf{J}_\text{cyt} \cdot \mathbf{n} \big|_{\mathcal{S}} = \mathcal{R}_\text{boundary}(\mathbf{m}, \mathbf{c}|_{\mathcal{S}})}$.
It closes Eqs.~\eqref{eq:rds} and ensures \emph{local mass conservation} by equating the diffusive cytosolic flux onto the membrane (outward-pointing normal vector $\mathbf{n}$) with the net reactive flux due to protein attachment and detachment between cytosol and membrane.
The reaction term ${\mathcal{R}_\text{boundary}(\mathbf{m}, \mathbf{c}|_{\mathcal{S}})}$ contains all the attachment and detachment reactions also included in ${\cal R}_\text{mem}(\mathbf{m},\mathbf{c}|_{\mathcal{S}})$, but ${\cal R}_\text{mem}(\mathbf{m},\mathbf{c}|_{\mathcal{S}})$ additionally contains reactions solely involving membrane-bound components.

\subsection{Two-component mass-conserving reaction--diffusion systems}
\label{sec:2cMcRD_phase-space_analysis}

Although a full understanding of protein pattern formation requires physiological detail, key principles emerge from minimal models. 
These typically describe a single protein species cycling between cytosolic and membrane-bound states via biochemical reactions. 
By focusing on lateral redistribution along a flat membrane and neglecting vertical cytosolic gradients, the dynamics reduce to: 
\begin{subequations}
\begin{align}  
    \partial_t m(\mathbf{x},t) &= D_m \nabla^2 m + f(m,c) \, ,\\ 
    \partial_t c(\mathbf{x},t) &= D_c \nabla^2 c - f(m,c)  \, .
\end{align}
\end{subequations}
In this simplification, we have chosen both densities $m$ and $c$ as area densities with the same units.
A common form of the reaction term, $f(m,c) = a(m) \, c - d(m) \,  m$, captures self-recruitment and enzymatic detachment via density-dependent rates $a(m)$ and $d(m)$.
Two-component McRD systems have become valuable conceptual models for cell polarity~\cite{Edelstein-Keshet.etal2013, Halatek.etal2018, Frey.Brauns2022}.
Inspired by Rho GTPase networks, early models showed that nonlinear feedback in membrane–cytosol cycling can drive symmetry breaking via diffusion-driven lateral instabilities~\cite{Otsuji.etal2007, Ishihara.etal2007, Goryachev.Pokhilko2008}.
In parallel, the wave-pinning mechanism demonstrated how fast cytosolic diffusion, together with mass conservation, can halt a propagating front and stabilize a polarized state~\cite{Mori.etal2008}.
More recently, two-compartment McRD systems have been framed within a geometric phase-space approach that reveals how local equilibria, mass redistribution, and diffusive coupling govern the emergence of mesoscale patterns~\cite{Halatek.Frey2018, Brauns.etal2020, Frey.Brauns2022}.

\begin{figure*}[thb]
\centering
\includegraphics[]{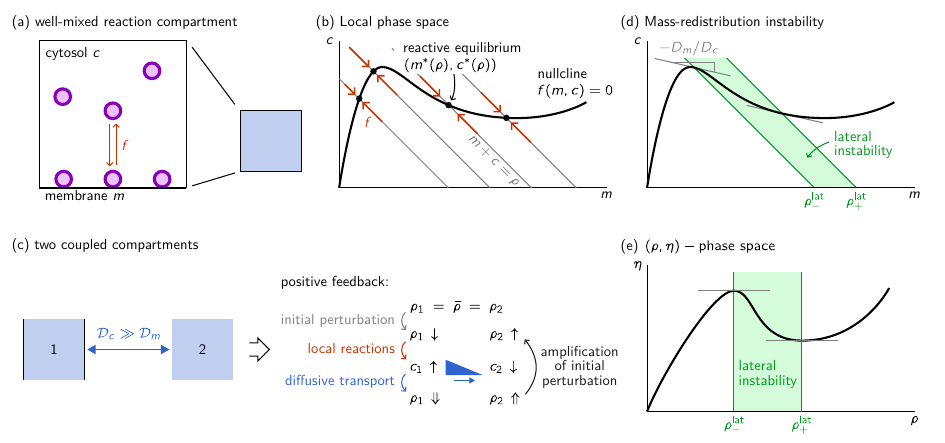}
\caption{
\textbf{Mass-redistribution instability.}
(a) In a single well-mixed compartment, the reaction term $f$ (red arrows) converts the cytosol ($c$) and membrane ($m$) densities into each other until a reaction equilibrium is reached.
(b) During the reactive relaxation in the single compartment, the total density ${\rho=m+c}$ remains constant, restricting the reactive flow to diagonals in the local phase space (grey lines).
The family of reactive equilibria (black dots) for different total densities $n$ form the nullcline (black).
(c) The diffusive coupling of two compartments via cytosolic exchange (neglecting membrane diffusion) induces a lateral instability if the nullcline slope is negative.
The underlying cause is a positive feedback resulting in self-amplifying mass transport between the two compartments.
(d) Including membrane diffusion, the mass-redistribution instability occurs for total densities ${\rho\in [\rho_-^\mathrm{lat},\rho_+^\mathrm{lat}]}$ at which the slope of the nullcline is smaller than $-D_m/D_c$ (green-shaded region).
(e) The local phase space can also be analyzed using the total density $n$ and the mass-redistribution potential $\eta$ as coordinates.
Mass redistribution induces a lateral instability if the nullcline in the $(\rho,\eta)$-phase space has a negative slope, also for finite membrane diffusion.
}
\label{fig:McRD_Phase-Space}
\end{figure*}

\subsubsection{Phase-space analysis}
\label{sec:phase_space_analysis}

To clarify the geometric structure underlying mass-conserving dynamics, it is helpful to coarse-grain space into two compartments—representing, for example, the polar zones of an E.~coli cell~\cite{Brauns.etal2021}. This minimal model retains essential spatial coupling while enabling a stepwise phase-space analysis: first of the local reactive dynamics in isolated compartments, then of their diffusive coupling.

We begin with the reactive dynamics in a single well-mixed compartment (Fig.~\ref{fig:McRD_Phase-Space}a), which can be analyzed geometrically in the $(m,c)$-phase space (Fig.~\ref{fig:McRD_Phase-Space}b). 
Local mass conservation constrains the dynamics to lines of constant total density ${m + c = \rho}$, called \emph{reactive subspaces}. 
Within each subspace, the system relaxes toward a \emph{reactive equilibrium} at the intersection with the \emph{reactive nullcline} $f(m, c) = 0$, where attachment and detachment balance.
For simplicity, we focus on monostable kinetics, where each total density $\rho$ corresponds to a unique stable reactive equilibrium that governs the dynamics within its subspace; see Ref.~\cite{Brauns.etal2020} for extensions to multistable systems.
To examine how spatial coupling affects stability, we consider two such compartments (Fig.~\ref{fig:McRD_Phase-Space}c), with local concentrations $m_i$, $c_i$, and total densities $\rho_i = m_i + c_i$, for $i = 1, 2$. Assuming cytosolic diffusion dominates exchange and membrane diffusion is negligible, the total densities evolve according to
\begin{equation}
\partial_t \rho_1 = \mathcal{D}_c \, (c_2 - c_1), \quad \partial_t \rho_2 = -\partial_t \rho_1,
\end{equation}
where $\mathcal{D}_c$ is an effective cytosolic exchange rate. 

To analyze the onset of instability, consider a small perturbation to a homogeneous state with equal densities ${\bar{\rho} = (\rho_1 + \rho_2)/2}$ in both compartments, such that ${\rho_{1,2} = \bar{\rho} \mp \Delta \rho}$ (Fig.~\ref{fig:McRD_Phase-Space}c).  
Assuming a separation of timescales with fast reaction kinetics and slow cytosolic exchange, each compartment rapidly relaxes to its local reactive equilibrium before appreciable mass transfer occurs. 
This relaxation onto the chemical equilibrium densities establishes a cytosolic concentration gradient because the total-density difference $2\Delta\rho$ between the two compartments implies different (cytosolic) densities which balance the attachment and detachment reactions (Fig.~\ref{fig:McRD_Phase-Space}b).
The sign of the cytosolic density difference is determined by the slope of the reactive nullcline, $\partial_m c^*(m)$.  
If the nullcline slope is negative, the cytosolic concentration decreases with increasing total density because the balance between attachment and detachment is shifted toward stronger attachment, implying ${c_1 > c_2}$. 
Diffusive transport in the cytosol then drives proteins from the left to the right compartment, further increasing $\rho_2$ and amplifying the initial perturbation (Fig.~\ref{fig:McRD_Phase-Space}c).
This positive feedback constitutes a \emph{mass-redistribution instability}, driven by self-amplifying diffusive exchange.  
By contrast, a positive nullcline slope implies ${c_1 < c_2}$, leading to a restoring flux that suppresses the perturbation and returns the system to homogeneity.

\subsubsection{Mass-redistribution potential}
\label{sec:mass_redistribution_potential}

The simplified compartmental treatment can be extended to spatially continuous systems, fully accounting for both membrane and cytosolic diffusion. 
The essential structural feature of McRD systems is that the dynamics of the total density ${\rho = m + c}$ can be rewritten as a continuity equation,
\begin{equation}
    \partial_t \rho (\mathbf{x},t)
    = 
    D_c \, \nabla^2 \eta
    \,,
    \label{eq:2cMcRD_continuity}
\end{equation}
with the \emph{mass-redistribution potential} defined as~\cite{Otsuji.etal2007,Ishihara.etal2007,Forte.etal2019,Brauns.etal2020}
\begin{equation}
    \eta (\mathbf{x},t) := c(\mathbf{x},t) + \frac{D_m}{D_c} \, m(\mathbf{x},t)
    \,.
    \label{eq:eta_def}
\end{equation}
This implies that protein flux is given by ${\mathbf{J} = -D_c \nabla \eta}$, so the mass-redistribution potential $\eta$ governs diffusive transport analogously to a chemical potential in near-equilibrium systems, such as in the Cahn--Hilliard model~\cite{Cahn.Hilliard1958} for phase separation. 
However, unlike chemical potentials derived from free energy functionals, $\eta$ is not variational; 
it evolves dynamically as part of the reaction--diffusion system:
\begin{align}
    \partial_t \eta(\mathbf{x},t) 
    &= (D_m + D_c)\, \nabla^2 \eta
    - D_m \, \nabla^2 \rho 
    \nonumber \\
    &\quad - (1 - d) \, f \big(m(\rho,\eta),\, c(\rho,\eta)\big)
    \,,
    \label{eq:McRD_eta_b}
\end{align}
where ${d = D_m / D_c <1}$.
Assuming that local reaction kinetics are fast relative to diffusive transport, we apply the \textit{local quasi-steady-state} (LQSS) approximation $f(m,c) = 0$.  
This defines local reactive equilibria, with concentrations approximated by $m(\mathbf{x},t) \approx m^*(\rho(\mathbf{x},t))$ and $c(\mathbf{x},t) \approx c^*(\rho(\mathbf{x},t))$.  
Substituting into the continuity equation, Eq.~\ref{eq:2cMcRD_continuity}, the dynamics reduce to a closed nonlinear diffusion equation for the total density:
\begin{align}
    \partial_t \rho(\mathbf{x},t) 
    = 
    D_c \, \nabla \cdot \left[ \partial_\rho \eta^*(\rho) \, \nabla \rho \right]
    \,,
\end{align}
which describes protein transport driven by $-\nabla \eta^*$, resulting in an effective diffusion coefficient ${D_\mathrm{eff}(\rho) = D_c \, \partial_\rho \eta^*(\rho)}$.
Pattern formation sets in when the uniform steady state $\rho(\mathbf{x}) = \rho_\mathrm{hss}$ becomes unstable due to negative effective diffusion (Fig.~\ref{fig:McRD_Phase-Space}(d,e)):
\begin{equation}\label{eq:slope-criterion}
    \partial_\rho \eta^*(\rho) < 0\,.
\end{equation}
In this regime, small density perturbations are amplified as mass flows from regions of lower to higher $\rho$, reinforcing inhomogeneities.
This is the hallmark of pattern formation driven by \textit{mass redistribution}.  
In the limiting case ${D_m = 0}$, this reduces to ${\partial_m c^*(m) < 0}$, consistent with the heuristic analysis in Sec.~\ref{sec:phase_space_analysis}.  
Importantly, the slope criterion (Eq.~\ref{eq:slope-criterion}) remains valid even beyond the LQSS approximation:  
while the exact timescale of the instability may differ, the sign of $\partial_\rho \eta^*(\rho)$ still determines whether local reactions increase or decrease $\eta$, and thus whether diffusive transport reinforces or counteracts density perturbations~\cite{Brauns.etal2020}.

\section{THE MOLECULAR BASIS OF PATTERN FORMATION IN THE \textit{ESCHERICHIA COLI} MIN PROTEIN SYSTEM}
\label{sec:Min_system}

We now turn to specific cellular systems and ask: \textit{How molecular is molecular?} That is, which features of protein interaction networks are essential to explain the emergent properties of self-organized patterns?
A common view in theoretical physics holds that coarse-grained models suffice to capture core mechanisms.
And indeed, many features—such as symmetry breaking and geometry sensing—can be understood at a phenomenological level, without molecular detail. 
Yet the \textit{E. coli} Min system demonstrates that specific molecular interactions are indispensable for explaining the robustness, diversity, adaptability, and functional integration of intracellular patterns.
In this section, we discuss how increasingly detailed models, starting from the core reaction cycle and extending to conformational switching and membrane interactions, account for the dynamic behavior of the Min system.

\subsection{Core biochemical cycle and minimal reaction–diffusion models}  
\label{sec:min-system-overview}

Symmetric division in \emph{E. coli} requires precise midcell placement of the FtsZ-based Z-ring. This positioning is guided by the Min system—a self-organizing protein network (MinC, MinD, MinE) that prevents Z-ring assembly near the poles~\cite{Lutkenhaus2007}.

\subsubsection{ATPase-driven reaction cycle}  
\label{sec:min-atpase-cycle}

The dynamics of the Min system are governed by a cyclical, ATPase-driven reaction network that controls membrane attachment and detachment (Fig.~\ref{fig:introduction}b).  
In its ATP-bound form, MinD dimerizes and binds cooperatively to the  membrane~\cite{Hu.etal2002, Suefuji.etal2002, Szeto.etal2002, Szeto.etal2003, Hu.Lutkenhaus2003, Hu.etal2003, Lackner.etal2003, Mileykovskaya.etal2003, Taghbalout.etal2006}.  
Membrane-bound MinD--ATP recruits both the division inhibitor MinC and its ATPase-activating partner MinE~\cite{deBoer.etal1989, Raskin.DeBoer1997, Hu.Lutkenhaus1999, Raskin.DeBoer1999a, Hu.etal2002, Suefuji.etal2002, Lackner.etal2003, Hu.etal2003, Wu.etal2011}.  
MinE stimulates ATP hydrolysis by MinD, triggering release of MinDE complexes into the cytosol as MinD--ADP and MinE~\cite{Raskin.deBoer1999, Rowland.etal2000, Hu.Lutkenhaus2001, Hu.etal2002, Suefuji.etal2002, Lackner.etal2003, Hu.Lutkenhaus2003}.  
MinD is then reactivated via nucleotide exchange~\cite{Hu.etal2003}, closing the cycle.  
Since cytosolic and membrane-bound proteins diffuse at different rates, this reaction cycle is inherently coupled to spatial protein redistribution.

\subsubsection{From activator–inhibitor models to mass-conserving dynamics}
\label{sec:activator-inhibitor_McRD}

The ATPase cycle described above lacks a mechanism for local MinD accumulation and cannot, by itself, explain spontaneous symmetry breaking and spatiotemporal pattern formation.
Different pattern-forming feedback mechanisms were thus discussed in theoretical models \cite{Meinhardt.DeBoer2001,Howard.etal2001,Kruse2002}.
A key insight by Howard et al.\ \cite{Howard.etal2001} and Kruse \cite{Kruse2002} was the formulation of McRD systems based on the observation that Min oscillations persist after inhibiting protein synthesis~\cite{Raskin.deBoer1999}.
This observation implied that patterns arise from protein redistribution rather than synthesis–degradation cycles as assumed in classical activator--inhibitor models~\cite{Meinhardt.DeBoer2001}.  
While models implement the core cycle of MinD membrane binding and MinE-triggered detachment, they differ in the nature of the feedback mechanisms and in their assumptions on MinE recruitment. 
Howard et al.~\cite{Howard.etal2001} showed that mutual suppression of MinD and MinE binding rates can generate oscillations, but the model did not support MinE recruitment by membrane-bound MinD~\cite{Raskin.DeBoer1997, Hale.etal2001, Suefuji.etal2002, Hu.etal2002}. 
To account for this, nonlinear feedback via MinD self-recruitment or lateral aggregation were introduced~\cite{Meinhardt.DeBoer2001,Kruse2002}.  
Later MinD oligomerization has been observed \emph{in vitro}~\cite{Hu.etal2002,Suefuji.etal2002,Miyagi.etal2018,Heermann.etal2021} and suggested \emph{in vivo}~\cite{Shih.etal2003}, providing mechanistic support for such feedback.  
Particle-based models further showed that MinD filamentation can drive oscillations~\cite{Tostevin.Howard2005,Pavin.etal2006}.
While these models introduced the role of cooperative MinD interactions, they relied on phenomenological assumptions—such as broken mass conservation~\cite{Meinhardt.DeBoer2001}, suppression of MinE recruitment at high MinD density~\cite{Meinhardt.DeBoer2001,Kruse2002}, or unrealistically high membrane mobility~\cite{Kruse2002}—not supported experimentally~\cite{Raskin.deBoer1999,Huang.etal2003}.

\subsubsection{Core feedback mechanisms driving Min pattern formation}  
\label{sec:min-feedback-mechanisms} 

\begin{figure*}
\centering
\includegraphics[]{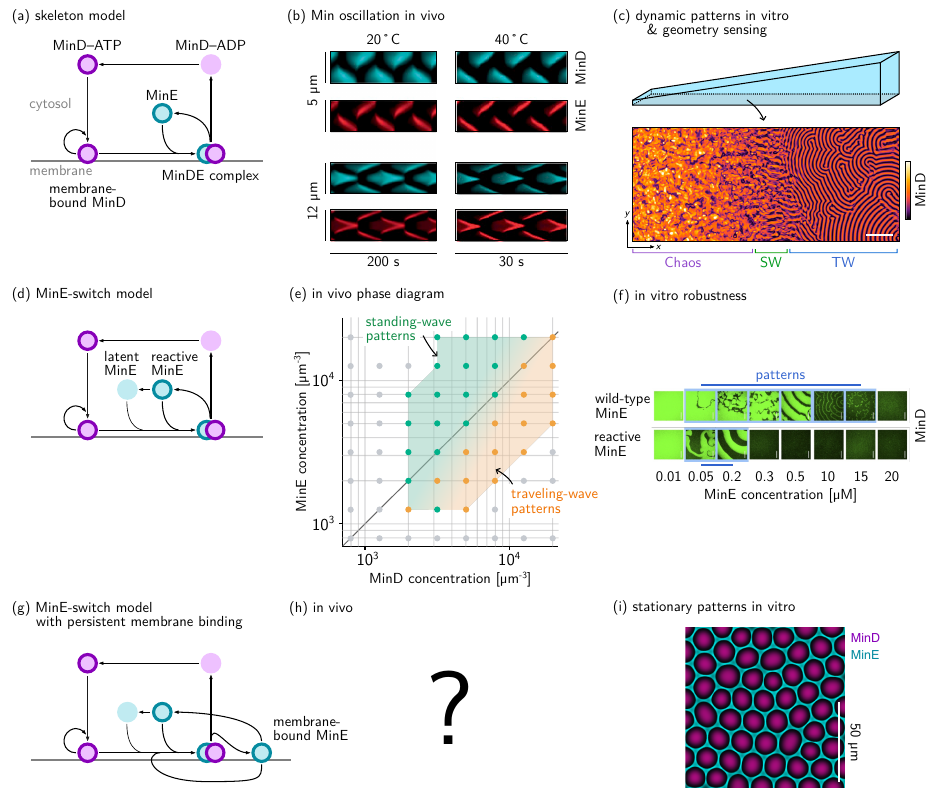}
\caption{
\textbf{Modules of the Min interaction network}.
(a) The skeleton model captures the core ATPase cycle of MinD.
MinD-ATP attaches to the membrane and recruits itself.
MinE is recruited as well and stimulates the ATPase activity of MinD, leading to the detachment of both proteins.
(b) The skeleton model results in pole-to-pole oscillations in \emph{in vivo} geometry (top).
Simulation in filamentous cells uncovers the intrinsic wavelength of the pattern, resulting in a standing-wave pattern (bottom).
The skeleton model captures the temperature dependence of the oscillations. Adapted from Ref.~\cite{Halatek.Frey2012}.
(c) In \emph{in vitro} geometries, the skeleton model captures chaotic, standing-wave, traveling-wave pattern, sensing the geometry via variations in the bulk--surface ratio. Adapted from Ref.~\cite{Wurthner.etal2022}.
(d) The MinE-switch model extends the skeleton model by a conformational switch between reactive and latent MinE states in the cytosol.
Reactive MinE is recruited much more quickly by membrane-bound MinD than MinE in the latent conformation.
(e) The MinE-switch model well describes the phase diagram of Min pattern formation in filamentous \textit{E.\ coli} bacteria.
It captures the robustness of pattern formation in protein concentrations, the pattern types, as well as the wavelengths and oscillation periods. Adapted from Ref.~\cite{Ren.etal2025}.
(f) \emph{In vitro}, mutation of the MinE protein showed that impairing its conformational switch (reactive MinE) strongly reduces the range of pattern formation, in accordance with the prediction by the MinE-switch model. Adapted from Ref.~\cite{Denk.etal2018}.
(g) MinE may persistently bind to the membrane via its membrane-targeting sequence.
(h) The relevance of MinE membrane binding for the Min oscillation \emph{in vivo} remains to be clarified by combined experiments and theory.
(i) MinE membrane-binding allows for the formation of stationary patterns observed with wild-type proteins \emph{in vitro}.
Scale bars in panels (c,f,i): $\SI{50}{\micro m}$. Adapted from Ref.~\cite{Weyer.etal2024b}.
}
\label{fig:switch-pmb-models}
\end{figure*}

Building on these earlier models, Huang et al.~\cite{Huang.etal2003} introduced a model solely based on biochemically suggested interactions.
This framework reproduced a wide range of intracellular Min patterns, including standing waves and MinE-ring formation.
However, the model requires a nucleotide-exchange rate lower than the experimentally suggested rate~\cite{Meacci.etal2006} and does not allow for sufficient parameter variations to explain pattern formation under varying temperature~\cite{Touhami.etal2006}.
To clarify the minimal set of experimentally supported interactions, the reaction network was systematically reanalyzed~\cite{Halatek.Frey2012}, leading to a simplified formulation now commonly referred to as the \emph{skeleton model}~\cite{Fange.Elf2006, Halatek.Frey2012}.  
This model retains MinD--ATP binding with self-recruitment, MinDE complex formation, MinE-stimulated detachment, and cytosolic nucleotide exchange—all within a McRD framework (Fig.~\ref{fig:switch-pmb-models}a--c).  
It accurately captures pole-to-pole oscillations and standing-wave patterns~\cite{Halatek.Frey2012}, and explains their dependence on system parameters such as protein copy number as well as on cell geometry~\cite{Halatek.Frey2012, Wu.etal2015, Wu.etal2016}. Stochastic effects, which become relevant at low copy numbers, are also well reproduced~\cite{Fange.Elf2006}.
In addition, the model accounts for qualitative transitions between distinct dynamic states induced by geometry~\cite{Thalmeier.etal2016, Halatek.Frey2018, Brauns.etal2021a, Wurthner.etal2022} and MinE mutations in vitro~\cite{Glock.etal2019}. 
This model thus offers a robust baseline for theoretical and experimental investigations of Min protein dynamics.

\subsection{Functional modules of the Min system}
\label{sec:min-functional-modules}

Beyond the core reaction cycle, a set of molecular mechanisms modulate the robustness and diversity of Min patterns. 
These function as regulatory modules: conformational switching buffers MinE activity, persistent membrane binding tunes pattern morphology, and their integration captures both \emph{in vivo} and \emph{in vitro} behavior. 

\subsubsection{Conformational switching of MinE: a robustness mechanism}
\label{sec:minE-switch}

The skeleton model supports pattern formation only within a narrow range of MinE/MinD concentration ratios—typically below one~\cite{Halatek.Frey2012}. 
This is at odds with experimental findings, initially from \emph{in vitro} reconstitutions~\cite{Denk.etal2018}, which show robust pattern formation across a broad range of MinD and MinE concentrations. 
This discrepancy pointed to the need for an additional regulatory mechanism and led to the introduction of the \textit{MinE-switch model} (Fig.~\ref{fig:switch-pmb-models}(d--f)~\cite{Denk.etal2018}. 
Motivated by experimental observations that MinE adapts both a reactive and a latent MinE conformation in which the MinD-binding interface is buried~\cite{Ghasriani.etal2010,Park.etal2011,Park.etal2017,Ayed.etal2017}, the MinE-switch model includes these two conformational MinE states~\cite{Denk.etal2018}.
While MinE quickly transitions into the latent conformation in the cytosol, sensing of MinD leads to the exposure of the binding site~\cite{Park.etal2011}, allowing for subsequent MinD binding.
This intermediate step is modeled by a strongly reduced recruitment rate of latent compared to reactive MinE. 

Theoretical analysis has shown that this switching behavior acts as a buffering mechanism~\cite{Denk.etal2018}.
The MinD-dependent switch from the latent into the reactive conformation ensures that only as much MinE is activated as is required to stimulate MinD hydrolysis.
As a result, at low MinE concentrations, most MinE is in the reactive form, while at higher concentrations, the excess MinE accumulates in the latent state~\cite{Ren.etal2025}.
This dynamic regulation extends the range of MinE/MinD ratios that support pattern formation.
The conformational switch has since been shown to explain the response of Min patterns to bulk flow~\cite{Wigbers.etal2020a,Meindlhumer.etal2023}. 
It also acts as a robustness module for pattern formation \emph{in vivo}~\cite{Ren.etal2025}, and furthermore, accounts for the emergence of different pattern types, including standing and traveling waves, as well as the observed dependence of wavelength and period on protein levels~\cite{Ren.etal2025,Vashistha.etal2023}. 
These results establish the MinE conformational switch as a key molecular feature that supports the adaptability and robustness of the Min system under physiological and reconstituted conditions.

\subsubsection{Persistent membrane binding of MinE: enhancing pattern diversity}
\label{sec:persistent_membrane-binding}

\emph{In vitro} experiments have revealed a rich diversity of Min protein patterns, including traveling and standing waves, bursts, mushrooms, and labyrinthine structures~\cite{Ivanov.Mizuuchi2010,Vecchiarelli.etal2016, Mizuuchi.Vecchiarelli2018, Glock.etal2019a, Glock.etal2019}. 
Strikingly, the position of the His-tag used for MinE purification—either at the membrane-targeting N-terminus~\cite{Hsieh.etal2010, Park.etal2011} or at the C-terminus—induces marked changes in pattern morphology, shifting the patterns from traveling waves to stationary structures such as spots, amoebas, and meshes~\cite{Glock.etal2019a}.
These phenotypes are not fully captured by either the skeleton~\cite{Halatek.Frey2012, Fange.Elf2006} or MinE-switch~\cite{Denk.etal2018} models.
The sensitivity to His-tag position may reflect effects on MinE’s membrane-binding behavior, although MinD-independent MinE membrane binding appears unaffected~\cite{Glock.etal2019a}.
While not essential for robust pattern formation and the qualitative description of patterns formed with His-MinE~\cite{Denk.etal2018}, MinE’s transient association with the membrane increases the pattern wavelength~\cite{Kretschmer.etal2017, Denk.etal2018} and influences pattern type for MinE-His~\cite{Vecchiarelli.etal2016, Glock.etal2019a}.
Thus, one proposed model refinement is \textit{persistent membrane binding} (PMB) of MinE. In this scenario, MinE remains transiently bound to the membrane after triggering ATP hydrolysis and MinD detachment~\cite{Loose.etal2011, Schweizer.etal2012, Bonny.etal2013,Wettmann.etal2018}. 
This leads to the formation of a local, membrane-associated pool of reactive MinE, capable of reengaging with MinD.
\emph{In vitro} experiments have shown that MinE lags behind MinD in traveling waves, supporting transient MinE membrane binding independent of MinD~\cite{Loose.etal2011, Vecchiarelli.etal2016}.
Indeed, including PMB in the skeleton model allowed to model labyrinthine patterns \cite{Fu.etal2023}, and inclusion in the MinE-switch model allows the robust formation of mesh patterns \cite{Weyer.etal2024b} (Fig.~\ref{fig:switch-pmb-models}c).
Future studies should analyze the interplay of the MinE switch and persistent membrane binding and aim for a unifying explanation of pattern formation \emph{in vivo} and \emph{in vitro}.

\section{INTERFACE DYNAMICS AND MESOSCOPIC LAWS}
\label{sec:interface-dynamics-mesoscopic-laws}

\begin{figure*}[htbp]
\centering
\includegraphics[width=0.8\linewidth]{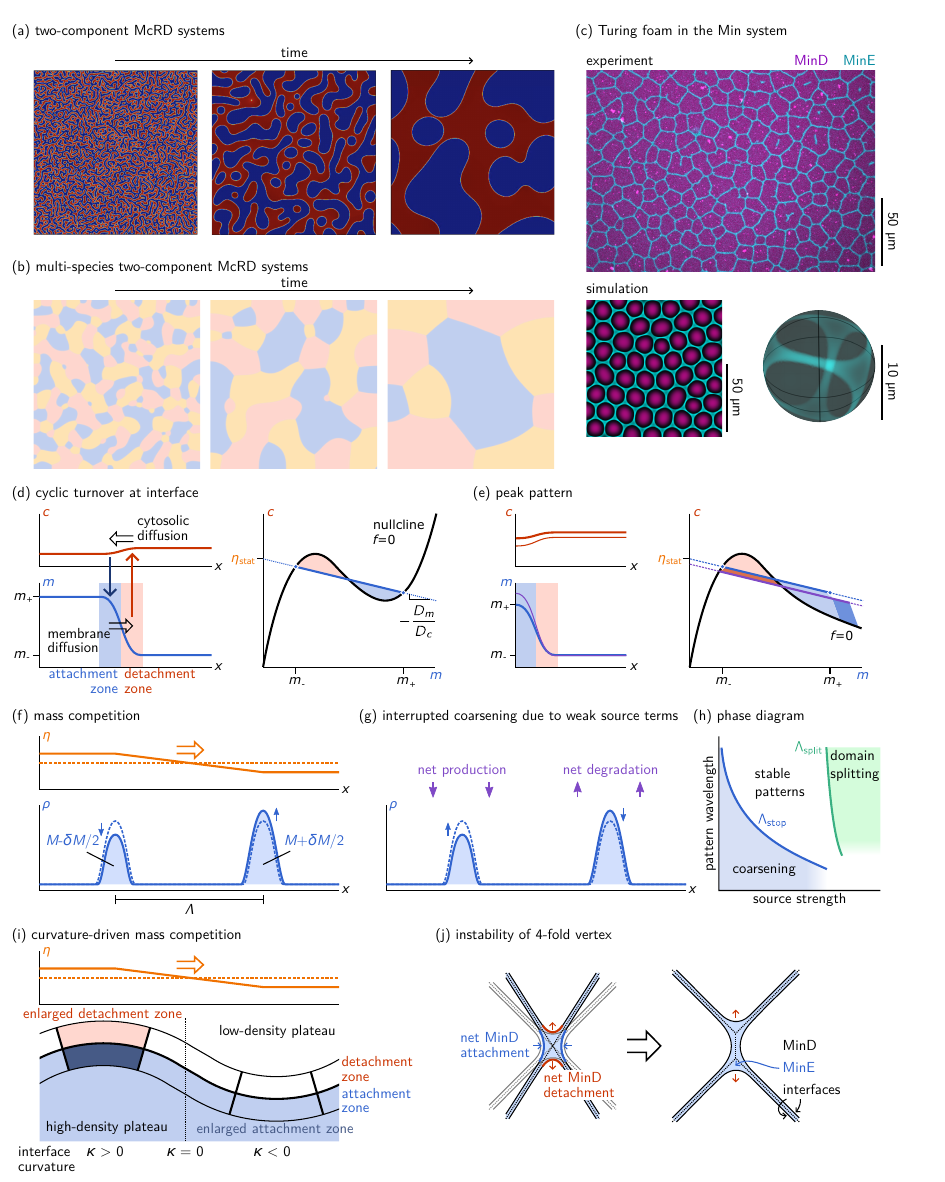}
\caption{
\textbf{Interfaces in McRD systems.}
(a) Two-component McRD systems form nonlinear patterns featuring interfaces between high- and low-density plateaus (or peaks, see (e)) that undergo coarsening.
(b) In three-species McRD systems, antagonistic reactions generate distinct membrane domains (blue, red, yellow).
(c) The Min system forms stationary mesh patterns, resembling 2D liquid foams \emph{in vitro}~\cite{Glock.etal2019a} and in simulations~\cite{Weyer.etal2024b}, where MinE-rich branches (cyan) separate MinD-rich domains (magenta). 
On a sphere, MinE forms polyhedral meshes (cyan); MinD not shown. Experimental data from Ref.~\cite{Glock.etal2019a}.
(d) Interfaces arise from balanced attachment and detachment zones, forming non-equilibrium steady states. 
In local $(m,c)$ phase space (right), the pattern lies on the flux-balance subspace (dashed), with plateau densities $m_\pm$ at its outer intersections with the nullcline. Balance of attachment (blue) and detachment (red) areas determines $\eta_\mathrm{stat}$.
(e) Peak patterns occur when high-density plateaus are not reached. 
In phase space, the pattern ends before the right-most intersection. 
Larger peaks (thin, purple) correspond to lower stationary mass-redistribution potential (area comparison).
(f) This potential dependence induces a mass-competition instability: 
small mass differences between peaks self-amplify (blue arrows).
(g) Weakly broken mass conservation introduces net production (smaller peak) and degradation (larger peak) that counteract this instability (blue arrows).
(h) As a result, coarsening halts above a wavelength $\Lambda_\mathrm{stop}$, set by the source strength; at larger wavelengths, domain splitting occurs above $\Lambda_\mathrm{split}$.
(i) In 2D, spatial separation of attachment and detachment zones leads to curvature-dependent turnover, driving interface straightening (orange arrow).
(j) This curvature dependence destabilizes fourfold vertices: increased detachment in curved regions (red) vs. flatter ones (blue) causes vertex splitting into pairs of triple junctions.
Panels b,c,i,j adapted from Ref.~\cite{Weyer.etal2024b}.
}
\label{fig:McRD_Interfaces}
\end{figure*}

The Min system illustrates how McRD systems generate large-amplitude patterns following an initial linear instability. 
In the nonlinear regime, quasi-stationary patterns emerge with sharp \emph{interfaces} separating membrane regions of distinct molecular composition and density (Fig.~\ref{fig:McRD_Interfaces}a--c). 
Interfaces are collective degrees of freedom, shaped by the coupling of interface geometry, diffusion, and reaction kinetics, and have long been used to describe the slow dynamics of nonlinear patterns. 
A paradigmatic case is the Schl\"ogl model of bistable kinetics, where domain boundaries encode front motion and coarsening~\cite{Schlogl1972}.
General interface-based theoretical approaches were developed for stationary~\cite{Petrich.Goldstein1994,Kerner.Osipov1994,Muratov.Osipov1996} and travelling~\cite{Hagberg.Meron1994,Pismen2006} patterns in two-component systems, motivated by the ferrocyanide-iodate-sulfite reaction--diffusion system~\cite{Lee.etal1993}. 
A growing body of work shows that also the large-scale dynamics of mass-conserving patterns can be effectively described by mesoscopic laws governing interface motion—laws that are largely independent of microscopic details and reflect robust, universal features of the underlying reaction–diffusion system.

\subsection{Stabilization of sharp interfaces via diffusive and reactive balance}
\label{sec:maintainig_interfaces}

Membrane diffusion tends to flatten density gradients.
In McRD systems, sharp density interfaces are maintained by localized attachment in the high-density region (\textit{attachment zone}) and detachment in the low-density region (\textit{detachment zone}) of the interface, which create a net reactive turnover along the interface (see Fig.~\ref{fig:McRD_Interfaces}d)~\cite{Goryachev.Pokhilko2008,Halatek.etal2018,Brauns.etal2020}. 
This turnover induces a diffusive counter-flux in the cytosol that balances membrane diffusion, stabilizing the interface as a non-equilibrium steady state driven by NTPase activity.
In two-component McRD systems, these principles provide the basis for constructing the interface profile in phase space~\cite{Brauns.etal2020}.
For a planar interface, balance between diffusive fluxes on the membrane and in the cytosol requires that the total flux perpendicular to the interface ${\sim\nabla_\perp\eta}$ vanishes (cf.~Eq.~\ref{eq:2cMcRD_continuity}).
Thus, a stationary interface is characterized by a constant stationary mass-redistribution potential, and, in the $(m,c)$-phase space, it lies on the \textit{flux-balance subspace} ${c+ d \, m = \eta_\mathrm{stat}}$ with $d=D_m/D_c$ (Fig.~\ref{fig:McRD_Interfaces}d).

In addition to diffusive flux balance, attachment and detachment must also balance at the stationary interface. In phase space, the shaded regions between the nullcline and the flux-balance subspace represent net attachment and detachment. 
Their balance determines the position of the flux-balance subspace, that is, the stationary value $\eta_\mathrm{stat}$, through an approximate area-matching condition (Fig.~\ref{fig:McRD_Interfaces}d). 
This construction is reminiscent of the classical Maxwell construction, though the underlying mechanism—based on local reactive turnover between membrane and cytosol—differs fundamentally from the thermodynamic criterion of equal osmotic pressures.
From this perspective, the wave-pinning mechanism~\cite{Mori.etal2008} appears as a limiting case of interface stabilization in mass-conserving systems with instantaneous cytosolic redistribution: the front becomes stationary when reactive turnover at the interface balances.
In the following, we show that the balance of attachment and detachment fully determines the dynamics of the pattern interfaces within a quasi-steady-state approximation for the local interface profile and mass-redistribution potential.

\subsection{Mass competition, interrupted coarsening, and pattern selection}

In two-component McRD systems, small perturbations of the homogeneous state are amplified by a mass-redistribution instability, giving rise to patterns with a characteristic wavelength determined by the fastest-growing linear mode~\cite{Halatek.Frey2018,Brauns.etal2020}. 
Over time, these patterns undergo coarsening~\cite{Otsuji.etal2007, Ishihara.etal2007, Goryachev.Pokhilko2008, Chiou.etal2018, Jacobs.etal2019, Brauns.etal2021b, Weyer.etal2023, Goryachev.Leda2020}: larger domains grow at the expense of smaller ones through mass exchange, ultimately leading to the dominance of a single domain (Fig.~\ref{fig:McRD_Interfaces}).

\subsubsection{Mass-competition instability and coarsening laws} 
\label{sec:mass_competition}

The coarsening process can be understood in terms of a \emph{mass-competition instability}—also known as \emph{winner-takes-all dynamics}~\cite{Goryachev.Pokhilko2008,Chiou.etal2018}—in which larger domains grow at the expense of smaller ones. This mechanism has been analyzed both mathematically~\cite{Ward2006, Wei.Winter2014, Kolokolnikov.etal2006, McKay.Kolokolnikov2012} and from a physical perspective~\cite{Otsuji.etal2007, Ishihara.etal2007, Brauns.etal2021b, Weyer.etal2023}.
Consider a regime in which well-separated, quasi-stationary patterns have formed and the dynamics are \emph{diffusion-limited}\footnote{Depending on whether diffusive transport or reactive conversion is rate-limiting~\cite{Weyer.etal2023}, the dynamics fall into either a diffusion-limited regime—analogous to Cahn–Hilliard dynamics~\cite{Cahn.Hilliard1958}—or a reaction-limited regime, resembling conserved Allen–Cahn systems~\cite{Rubinstein.Sternberg1992}.}. 
In one-dimensional (1d) systems, depending on the reaction kinetics, the emergent patterns are either mesa-shaped or peak-shaped (Fig.~\ref{fig:McRD_Interfaces}d,e)~\cite{Brauns.etal2020}.  
Focusing on peak patterns, we assume that each peak is in a (regional) quasi-steady state (QSS), such that the \emph{mass-redistribution potential} satisfies ${\eta = \eta_\text{stat}(M)}$, where ${M}$ is the total mass associated with a peak (Fig.~\ref{fig:McRD_Interfaces}e).
Differences in the peak masses result in gradients of the mass-redistribution potential that drive slow mass exchange (Fig.~\ref{fig:McRD_Interfaces}f).
For two neighboring peaks at a distance ${\Lambda}$, the dynamics of mass redistribution are governed by~\cite{Brauns.etal2021b, Weyer.etal2023}
\begin{equation}
    \partial_t \delta M \approx -\frac{2 D_c}{\Lambda} 
    \left. \frac{\partial \eta_{\text{stat}}}{\partial M} \right|_{M_0} \,
    \delta M,
\end{equation}
where ${\delta M}$ is the mass difference between the peaks.  
The total turnover balance implies that ${\partial_M \eta_\text{stat}(M) < 0}$ holds for stable peak and mesa patterns in two-component McRD systems (Fig.~\ref{fig:McRD_Interfaces}e)~\cite{Brauns.etal2021b}.
Thus, the symmetric pattern is unstable against the growth of one and collapse of the other peak and these systems generically undergo uninterrupted coarsening.

The wavelength dependence of the mass-competition rate determines the long-time coarsening law.
Peak patterns exhibit power-law coarsening, reflecting scaling behavior of the reaction term at large densities~\cite{Brauns.etal2021b, Weyer.etal2023}, whereas mesa patterns coarsen logarithmically~\cite{Brauns.etal2021b, Weyer.etal2023}, consistent with the coarsening of coexisting phases in 1d liquid mixtures~\cite{Kawasaki.Ohta1982,Kawasaki.Nagai1983,Nagai.Kawasaki1986,Alikakos.etal1991}.
A similar transition between logarithmic and power-law coarsening has been observed in thin films, where gravity-induced saturation of the droplet height leads to mesa-like droplet profiles~\cite{Glasner.Witelski2003,Glasner.Witelski2005,Gratton.Witelski2008}.

\subsubsection{Wavelength selection by interrupted coarsening and domain splitting} 

Uninterrupted coarsening of two-component McRD systems is rather surprising given that two-component systems without mass-conservation are the classical examples of Turing systems, which form patterns with an intrinsic wavelength, frequently approximated by the wavelength of the fastest-growing mode of the linear instability of the homogeneous steady state \cite{Turing1952,Maini.etal1997}.
These observations are reconciled by interrupted coarsening in systems with weakly broken mass conservation~\cite{Brauns.etal2021b, Weyer.etal2023}.
Similarly, coarsening is interrupted by coupling to a third component~\cite{Jacobs.etal2019,Chiou.etal2021,Gai.etal2020a}.
In systems with broken mass conservation, net production in low-density regions and degradation in high-density domains counteract the mass-competition instability (Fig.~\ref{fig:McRD_Interfaces}g).
This stabilizes periodic patterns above a threshold wavelength $\Lambda_\mathrm{stop}$, resulting in \emph{wavelength selection by interrupted coarsening}~\cite{Brauns.etal2021b, Weyer.etal2023}. 
At even larger wavelengths, domain plateaus become locally unstable and split, defining an upper bound $\Lambda_\mathrm{split}$ for stable pattern sizes. 
As a result, the phase diagram Fig.~\ref{fig:McRD_Interfaces}h unifies coarsening and wavelength selection of highly nonlinear patterns based on clear physical mechanisms. 
It also applies to chemically driven phase-separating binary mixtures~\cite{Glotzer.etal1995,Zwicker.etal2015,Li.Cates2020} and Keller--Segel models~\cite{Hillen.Painter2009,Kolokolnikov.etal2014,Meyer.etal2014,OByrne.Tailleur2020,Weyer.etal2025b,Weyer.etal2025c}.
Intriguingly, If the two thresholds ${\Lambda_\mathrm{stop,split}}$ lie close to each other, continued coarsening and splitting lead to a spatiotemporally chaotic dynamic steady state~\cite{Painter.Hillen2011,Ei.etal2014,Weyer.etal2025b}.

\subsection{Effective interfacial tension and interface laws in two-dimensional systems}

The mass-competition mechanism underlying coarsening in one-dimensional McRD systems remains central in two-dimensional systems, but the dynamics are further shaped by geometric effects, most notably \textit{interface curvature} in two-component systems and the \textit{geometry} and dynamics of \textit{junctions and vertices} in multi-species systems (Fig.~\ref{fig:McRD_Interfaces}b--c). 

\subsubsection{Effective interfacial tension induced by non-equilibrium fluxes}

Two-component systems have been observed numerically to minimize the length of pattern interfaces, resulting in a coarsening process~\cite{Tateno.Ishihara2021,Singh.etal2022}.
Mathematically, length minimization was derived in a two-component system in the limit of infinite cytosolic diffusion~\cite{Miller.etal2023}.
Moreover, specific forms of the reaction kinetics allow for a mathematical mapping of the two-component systems onto effective near-equilibrium phase-field models that are governed by an interfacial tension, and thus undergo coarsening \cite{Morita.Ogawa2010}.
Even beyond this specific mapping, pattern coarsening is driven by curvature-dependent mass transport~\cite{Brauns.etal2021b, Weyer.etal2024b}, consistent with a Gibbs--Thomson relation
and classical Lifshitz–Slyozov–Wagner (LSW) theory~\cite{Lifshitz.Slyozov1961,Wagner1961}.
These works suggest that patterns in two-component systems are governed by an \emph{emergent nonequilibrium interfacial tension} that results in curvature-driven interface motion analogous to binary liquid mixtures, and which also appears to underlie interface-length minimization observed in more complex protein patterning systems~\cite{Maree.etal2012, Gessele.etal2020}.

A recent theoretical analysis provides a mechanistic explanation for this emergent interfacial tension~\cite{Weyer.etal2024b}. 
At a flat interface, the value of the stationary mass-redistribution potential $\eta_\mathrm{stat}$ is set by the balance between attachment and detachment fluxes; see Sec.~\ref{sec:maintainig_interfaces} and Fig.~\ref{fig:McRD_Interfaces}d. 
Interface curvature stretches and compresses the attachment and detachment zones unevenly along the arc length due to their spatial separation within the interface, leading to an imbalance in the integrated fluxes (Fig.~\ref{fig:McRD_Interfaces}i).
This geometric asymmetry induces a shift in the stationary mass-redistribution potential at weakly curved interfaces that scales linearly with the local interface curvature $\kappa$ \cite{Brauns.etal2021b,Weyer.etal2024b}:
\begin{equation}
    \delta \eta_\text{stat}(\kappa) 
    \sim 
    \ell_\text{int} \, \kappa \, ,
    \label{eq:gibbs-thomson}
\end{equation}
where ${\ell_\text{int} \sim \sqrt{D_m \tau_r}}$ is the interface width, determined by membrane diffusivity and the timescale of reactive turnover.
This curvature-induced shift creates gradients in ${\eta}$ between differently curved regions of the pattern interface, which cause mass transport that drives its straightening. 
Because ${\eta}$ plays for mass redistribution and interface movement a role analogous to a chemical potential, the resulting curvature-driven interface motion mirrors the Gibbs–Thomson effect in equilibrium phase separation. 
However, the effect arises entirely from reaction–diffusion dynamics, without any underlying free energy or mechanical surface tension. 
The associated effective interfacial tension scales as ${\sigma \sim \ell_\text{int}}$, and is expected to apply broadly to multi-component systems, provided that feedback is mediated by slow-diffusing membrane-bound components and the interface remains monotonic~\cite{Weyer.etal2024b}.

\subsubsection{Non-equilibrium Neumann law at triple junctions}
\label{sec:neumann_law}

The curvature dependence of the mass-redistribution potential in McRD systems invites comparison of multi-species McRD systems with multi-component phase-separating mixtures, where different domain boundaries can meet at triple junctions. 
In  equilibrium mixtures, the meeting angles between interfaces are set by the classical \textit{Neumann law}, reflecting a force balance between surface tensions~\cite{DeGennes.etal2004}.
Remarkably, a closely analogous relation arises in McRD systems—despite the absence of mechanical forces—through the balance of attachment and detachment fluxes of all species.
When each protein species consists of a membrane-bound and a cytosolic component, and mutual antagonism enforces domain segregation, the interface angles at triple junctions are governed by a \textit{non-equilibrium Neumann law}~\cite{Weyer.etal2024b}:
\begin{align}
    \tilde{\bm\sigma}_{AB} 
    + 
    \tilde{\bm\sigma}_{AC} 
    + \tilde{\bm\sigma}_{BC} 
    = 
    \mathbf{T}_\text{core} 
    \, ,
\end{align}
where ${\tilde{\bm\sigma}_{ij}}$ is a vector parallel to the interface between domains ${i}$ and ${j}$ and a magnitude corresponding to an effective interfacial tension of this interface. 
${\textbf{T}_\text{core}}$ captures the excess reactive turnover at the junction due to cyclic non-equilibrium fluxes.
This non-equilibrium Neumann law links local biochemical interactions to global pattern geometry in multi-species McRD systems~\cite{Weyer.etal2024b}, analogously as in complex liquid mixtures~\cite{Mao.etal2019,Mao.etal2020}.

\subsubsection{Turing foams: non-equilibrium interface dynamics in the Min system}
Among the variety of stationary patterns observed in the \emph{in vitro} Min system~\cite{Ivanov.Mizuuchi2010,Vecchiarelli.etal2014,Mizuuchi.Vecchiarelli2018,Glock.etal2019a,Weyer.etalinpreparationc}, some exhibit morphologies that closely resemble two-dimensional liquid foams (Fig.~\ref{fig:McRD_Interfaces}c).
In these patterns, MinD-enriched membrane domains are separated by narrow MinE-enriched branches that predominantly meet at triple vertices.
Heuristically, we expect 4-fold and higher-order vertices to be unstable as perturbations of their symmetric configuration induces net MinD attachment and detachment around the vertex that leads to its splitting into separate triple vertices (Fig.~\ref{fig:McRD_Interfaces}j).
Moreover, the meeting angles at these vertices are tightly distributed around $120^\circ$, consistent with Plateau’s laws for liquid foams derived from surface-tension-driven surface minimization~\cite{Weaire.Hutzler2001}.
Simulations of the MinE-switch model (Sec.~\ref{sec:minE-switch}), including persistent membrane binding (Sec.~\ref{sec:persistent_membrane-binding}),  reproduce this foam-like geometry and vertex angle distribution~\cite{Weyer.etal2024b} (cf. Fig.~\ref{fig:McRD_Interfaces}c).

This analogy extends beyond morphology to dynamics. 
In the early stages of pattern evolution at small pattern wavelength, domain areas follow qualitatively a \textit{von Neumann–type law}: 
domains with fewer than six edges shrink, those with more than six grow, and six-edged domains remain stationary. 
This behavior is captured by the relation
\begin{equation}
    \partial_t A_n \sim (n - 6)
    \, ,
\end{equation}
where $A_n$ is the area of an $n$-sided domain. 
This classical result is rooted in the presence of an (effective) interfacial tension, the $120^\circ$ vertex angles, and the geometric constraint imposed by the Gauss–Bonnet theorem, which together determine the rate of area change in two-dimensional foams~\cite{Weaire.Hutzler2001,Weyer.etal2024b}. 
These findings demonstrate that Min mesh patterns obey mesoscopic laws analogous to those of liquid foams, motivating the term \emph{Turing foam} to underline that these patterns arise purely from the reaction--diffusion mechanism~\cite{Weyer.etal2024b}.

A key difference to liquid foams is that coarsening in Turing foams is interrupted: once domains exceed a critical size coarsening arrests, and at larger sizes domains undergo splitting through the growth of new MinE-rich branches~\cite{Weyer.etal2024b}.
This results in an intrinsic pattern wavelength, analogous to that of two-component McRD systems with weakly broken mass conservation.
The source terms in these simpler conceptual systems can be understood as a coupling to a third reservoir component, and it has been shown that also the coupling to a third diffusive component can interrupt coarsening and induce domain splitting~\cite{Jacobs.etal2019, Gai.etal2020a, Chiou.etal2021}.
An important open question is how the extra components in the Min system mechanistically give rise to interrupted coarsening and domain splitting.
Both the collapse of small and the splitting of large domains has been observed in mesh and amoeba patterns \cite{Ivanov.Mizuuchi2010,Weyer.etal2024b}, suggesting that these insights will enable a conceptual understanding of experimetally accessible, quasi-stationary protein patterns.

\section{CONCLUSION AND OUTLOOK}

This review has highlighted how mass-conserving reaction--diffusion systems provide a unifying framework for understanding protein-based pattern formation in living cells and reconstituted systems. 
Central to this framework are the concepts of mass redistribution, interface dynamics, and mesoscale laws such as curvature-driven coarsening and effective interfacial tension. 
The \textit{E.~coli} Min system has served as a paradigmatic example, demonstrating how spatial organization in cells can emerge from a small set of molecular interactions governed by universal physical principles.
Building on this foundation, several key directions emerge for advancing both the mechanistic and theoretical understanding  of protein-based pattern formation:

\medskip

\textbf{Mechanistic reduction of complex systems.}
Detailed biochemical models now quantitatively account for the dynamic behaviors of protein pattern formation across a wide range of experimental conditions. 
The next step is to reduce these models to minimal McRD frameworks that reveal core design principles.
Such reductions could clarify how how distinct molecular interactions give rise to the observed diversity of mesoscale patterns, an expose functional redundancies in the network~\cite{Brauns.etal2023}.

\medskip

\textbf{Coarse-graining molecular mechanisms.}
Although current models capture the emergent dynamics with quantitative accuracy, essential feedbacks—such as MinD self-recruitment—still lack a clear mechanistic derivation from microscopic processes like MinD oligomerization or higher-order MinE interactions. Bridging this gap between molecular detail and mesoscale dynamics remains a central theoretical task. 
Additional physical mechanisms, including diffusiophoretic coupling~\cite{Ramm.etal2021} and mechanochemical feedbacks such as curvature-sensitive protein recruitment~\cite{Goychuk.Frey2019}, may also play a role.
In view of future advances in protein engineering, such systematic connections  will be central to control and design pattern-forming feedbacks on the level of single proteins.

\medskip

\textbf{Geometry sensing and morphodynamics.}
Geometry sensing in McRD systems arises from the coupling between bulk and boundary dynamics~\cite{Thalmeier.etal2016, Halatek.Frey2018, Gessele.etal2020}.
Systematic dimensionality reduction~\cite{Burkart.etal2024} will be critical for an analytical understanding of this effect in complex geometries.
When proteins also deform the geometry, mechanochemical feedback can generate patterns even in minimal systems~\cite{Wurthner.etal2023}.
Reconstituted Min patterns have been shown to induce persistent vesicles motion~\cite{Fu.etal2023} and drive dynamic vesicle deformations~\cite{Litschel.etal2018, Christ.etal2021, Reverte-Lopez.etal2024}.
Moving beyond reconstitution, recent work in starfish oocytes has shown that intracellular protein patterns, coupled to the actin cortex, can sense cell geometry and drive large-scale mechanical deformations of the cortex~\cite{Liu.etal2025, Wigbers.etal2021, Wigbers.etal2020}.
These examples highlight how biochemical patterns can both sense and sculpt geometry, providing design principles for programmable morphodynamics.

\medskip

\textbf{Synthetic and reconstituted systems.} 
Beyond the Min system, spatial pattern formation has been reconstituted in other protein networks, including Rab5~\cite{Cezanne.etal2020, Bezeljak.etal2020} and a lipid kinase--phosphatase system~\cite{Hansen.etal2019}. Mitotic and Rho--actin waves have been observed in cell extracts~\cite{Chang.FerrellJr2013, Landino.etal2021}. Fully synthetic systems have implemented predator--prey oscillations, traveling waves, and bistable fronts using DNA circuits~\cite{Padirac.etal2013, Zadorin.etal2015}. Synthetic multicellular systems employing morphogen gradients~\cite{Stapornwongkul.etal2020}, SynNotch receptors~\cite{Morsut.etal2016}, or adhesion-based feedback~\cite{Toda.etal2018} now enable programmable tissue architectures. Together, these platforms support an emerging framework of synthetic developmental biology that bridges minimal biochemistry and engineered morphodynamics~\cite{McNamara.etal2023}. 

\medskip

\textbf{Theoretical challenges in nonequilibrium physics.}
The emergence of effective interface laws suggests that intracellular protein patterns obey universal mesoscale principles, largely independent of molecular detail. These insights point toward a broader theoretical framework for nonequilibrium pattern formation. 
As effective interfacial tensions are ubiquitous in active matter systems, and these form foam-like patterns both in theoretical models \cite{Maryshev.etal2020, Fausti.etal2021, DeLuca.etal2024a} and experiments \cite{Lemma.etal2022}, it will be fascinating to study foam formation and mesoscopic interface laws in non-equilibrium systems more broadly.

Pursuing these directions will not only deepen our understanding of biological self-organization but also advance a unified theory of nonequilibrium pattern formation that connects intracellular reaction--diffusion with active matter physics, and the design of synthetic systems.

\section*{ACKNOWLEDGMENTS}
This work was funded by the Deutsche Forschungsgemeinschaft (DFG, German Research Foundation) through the Excellence Cluster ORIGINS under Germany’s Excellence Strategy---EXC-2094---390783311, the European Union (ERC, CellGeom, project number 101097810), and the Chan-Zuckerberg Initiative (CZI).

\end{document}